\documentclass[a4paper,12pt,reqno]{amsart}
\usepackage{amssymb,amsmath,graphicx}

\newtheorem{lemma}{Lemma}

\newtheorem{comm}[lemma]{Comment}
\newtheorem{remm}[lemma]{Remark}
\newtheorem{exam}[lemma]{Example}

\newcommand{\alp}{\alpha}

\newcommand{\lam}{\lambda}

\newcommand{\CA}{\mathcal{A}}

\newcommand{\lap}{\Delta}

\newcommand{\R}{\mathbb{R}}
\newcommand{\C}{\mathbb{C}}

\newcommand{\Z}{\mathbb{Z}}

\newcommand{\ud}{{\rm d}}

\newcommand{\hull}{\mathrm{\hull}}

\newcommand{\dotp}[1]{\left< #1 \right>}

\newcommand{\Proof}[1][\!\!]{\textbf{\underline{Proof}.}\, #1}

\fontshape{n}

\title[Discreteness of the spectrum of the compactified...]
{discreteness of the spectrum of the compactified $D=11$
supermembrane with non-trivial winding}
\author[L. Boulton, M.P. Garc\'ia del Moral, A. Restuccia]{}

\date{$23^{th}$ July 2003}

\begin{document}
\

\vskip -1.5em \null

\vskip 1.5em

\maketitle

\centerline{\scshape L. Boulton$^1$, M. P. Garc\'\i a del
Moral$^2$, A. Restuccia$^2$}

\medskip

{\footnotesize \centerline{ $^{1}$Departamento de Matem\'aticas,
$^{2}$Departamento de F\'\i sica,} \centerline{Universidad Sim\'on
Bol\'\i var} \centerline{Apartado 89000, Caracas 1080-A,
Venezuela.} \centerline{emails: \texttt{lboulton@ma.usb.ve,
mgarcia@fis.usb.ve, arestu@usb.ve}}}

\medskip

\begin{abstract}
We analyze the Hamiltonian of the compactified \linebreak $D=11$ supermembrane
 with
non-trivial central charge in terms of the matrix
model constructed in \cite{1}. Our main result provides a rigorous proof
that the quantum Hamiltonian of the supersymmetric model has
compact resolvent and thus its spectrum consists of a discrete set
of eigenvalues with finite multiplicity.
\end{abstract}

\section{Introduction}
According to \cite{2}, the spectrum of the $SU(N)$
regularized supermembrane on $D=11$ Minkowski target space is
continuous and it consists on the whole interval $[0,\infty)$.
Although it was proven for a
regularized model, this result led to an interesting interpretation of the
supermem\-brane in terms of a multiparticle theory. It also showed
explicitly how the presence of supersymmetry may change completely
the spectrum of a bosonic discrete Hamiltonian over a compact
world volume. The proof was based on the existence of locally
singular configurations, which do not change the energy of the
system, and on the presence of supersymmetry.

The situation concerning the spectrum of the compactified
supermembrane is, in distinction, very different. Since the closed
but not exact modes present in the compactified case do not fit
into an $SU(N)$ formulation of the theory \cite{3}, the $SU(N)$
regularization of the compactified supermembrane seems not to be
possible. In \cite{3}, it was suggested  that the spectrum of
the compactified supermembrane should also be continuous due to
the existence of string-like spikes as in the non-compactified
case. The presence of those singular configurations, which do not
change the energy of the system, is a common property of all
p-branes \cite{4}. Recently, \cite{5}, it was shown that the
Hamiltonian formulation of the super M5-brane also contains
singular configurations even when they are neither present in the
known covariant formulation of the theory \cite{6} nor in \cite{7}.

In \cite{8} and \cite{9}, the compactified supermembrane on
$M_9\times S_1\times S_1$ was
formulated as a noncommutative gauge theory. The Hamiltonian was
given in terms of a noncommutative super Maxwell theory plus
the integral of the curvature of the noncommutative connection on
the world volume. In \cite{1}, it was explicitly shown that this
Hamiltonian allows the presence of string-like spikes in the
configuration space of the compactified supermembrane. This is in
agreement with the argument in \cite{3}. If however the theory
is restricted to a fixed central charge, by describing a sector
of the full compactified supermembrane, the Hamiltonian reduces
exactly to a noncommutative super Maxwell theory coupled to
seven scalar fields representing the transverse directions to the
supermembrane. In this case
it was shown in \cite{1} that there
are no string-like spikes in the configuration space. This result
is in agreement with the arguments in \cite{11}.

According to \cite{12}, the following properties for the bosonic part of this
Hamiltonian hold: it is bounded below 
and it becomes infinite at infinity
in every possible direction on the configuration space. In other words,
the potential is ``basin shaped''. This property ensures 
that the resolvent
of the bosonic Hamiltonian is compact and therefore its spectrum consists
of a set of isolated eigenvalues of finite multiplicity whose only
accumulation point is plus infinity. Furthermore, it is possible to find 
 upper bounds for the
asymptotic distribution of eigenvalues.

Based on two fundamental properties already discussed in \cite{12}, the
non existence of string-like spikes and the shape of the bosonic
potential, in the present paper we provide a rigorous proof of
the discreteness of the spectrum of the Hamiltonian of the
noncommutative super-Maxwell theory describing the compactified
$D=11$ supermembrane with fixed central charge. To be precise we
show that, as in the bosonic case, the resolvent is compact and
hence the spectrum consists of a discrete set of eigenvalues of
finite multiplicity. For this we provide a criterion (see lemma
\ref{t1}), that extends to the supersymmetric case, the well known
fact that  the spectrum of a Hamiltonian is discrete, when the
potential is bounded from below and unbounded above in all
directions (see \cite{13.1} and \cite{13.2}). The latter played a
fundamental role for the bosonic case discussed in \cite{12}. We state this
criterion in generic form and provide a self-contained proof,  so that 
it might be applied to
Hamiltonians whenever the bosonic potential forbids string-like
spikes.

To motivate the present analysis, we can comment on the relationship
 between our fixed central
charge model and the free model 
studied in \cite{2}, \cite{2.1} and \cite{3}. 
The winding supermembrane in the latter, can be regarded as a ``free''
model,
since it includes in the configuration 
space all possible wrappings,
$n=0,1,2,\ldots$,  on the compactified target space. In contrast, the
present work concerns only the sector of the full theory which corresponds
to a fixed central charge $n\not =0$. This case is also of physical relevance,
since  the fixed wrapping is a ``topological'' condition with phenomenological
applications. The change in the nature of the spectrum is formally analogous 
to the well known case
of the Laplacian acting on a domain of the Euclidean space: when the domain
is the whole space, the operator has pure continuous spectrum in contrast to
the case of bounded domains, where the operator has always pure discrete 
spectrum. We will discuss further on the comparison of the two models  
in section~2.


\section{The Hamiltonian of the double compactified supermembrane}
\label{s2}
Reference \cite{14} is devoted to study the
quantization of the compactified
supermembrane in an $M_{9}\times S^{1}\times S^{1}$ target space,
by finding an explicit expression, given in
terms of creation and annihilation operators,
of the quantum Hamiltonian in the semi-classical regime.
In particular,  there are no  vacuum energy
corrections to the mass formula, since the fermionic contribution
cancels the bosonic one.

The exact Hamiltonian was computed in \cite{9} (see also \cite{10}),
in terms of a symplectic noncommutative geometry. The symplectic
structure which gives rise
to the geometry, arises from the non-trivial central
charge originated in the wrapping of the supermembrane along the
compactified directions of the target space.  In order to make
more transparent the present exposition, we devote this section to review
the main
ideas which led to the construction of this Hamiltonian.

Let us start by considering the $D=11$ supermembrane Hamiltonian
in the light cone
gauge as in \cite{2}. In this model, the potential is given by
\begin{equation*} \label{n1}
    V(X)=\{X^m,X^n\}^2,\qquad \qquad m,n=1,\ldots,9
\end{equation*}
where $\{X^m,X^n\} \equiv (\epsilon^{ab}/\sqrt{W} )
\partial_a X^m\partial_b X^n$. Here, the scalar density $\sqrt{W}$ appears
in the formulation as consequence of the light cone gauge
fixing procedure.

Let $\Sigma$ be the spatial part of the world volume. We
always assume that $\Sigma$ is a compact Riemann surface of genus $g$.
If one of the target space spatial coordinates
is compactified on $S^1$, the natural winding condition is given by
\begin{equation} \label{n2}
   \oint _{c_j} \ud X = 2\pi m_j
\end{equation}
where $m_j$ are integers and $c_j$ is a basis of homology on
$\Sigma$. Analogously, when the target space has two compactified
directions on $S^1\times S^1$, we may consider $X_r$, $r=1,2,$ as
angular coordinates on each $S^1$. In order to have a well defined
map over $S^1$, we must impose as before the condition
\begin{equation} \label{mn1}
   \oint_{c_j} \ud X_r = 2\pi m_{jr}, \qquad r=1,2.
\end{equation}
Assume that the image of $\Sigma$ under $(X_1, X_2)$ describes
a torus. Then we should impose an additional constraint. If
$\Sigma$ itself
is a generic torus, where $w_1$ and $w_2$ denote the normalized
basis of homology on $\Sigma$, we have
\begin{equation*}
   \ud X_r=m_{1r}w_1+m_{2r}w_2+\ud a_r \qquad \qquad r=1,2
\end{equation*}
where $m_{jr},\,j=1,2,$ are the same integers introduced in \eqref{mn1} and
$\ud a_r$  are exact one-forms.
The requirement that the image of $\Sigma$ is a torus, 
may be interpreted as the
independence of the one-forms $m_{1r}w_1+m_{2r}w_2$ for $r=1,2,$ i.e.
\[
    \det (m_{jr})=n \not =0.
\]
This condition is equivalent to requiring
\begin{equation} \label{mn2}
 Z=\int_\Sigma (\ud X_r \wedge \ud X_s)\epsilon ^{rs}=2\pi n \not =0.
\end{equation}
The factor $2\pi$ corresponds to normalization of the area of $\Sigma$.
We remark that, since $Z$ becomes
\[
   \int _\Sigma g_1^{-1}\ud g_1\wedge g_2^{-1} \ud g_2
\]
where $G=(g_1,g_2)\in U(1)\times U(1)$, the integer $n$ is the winding number
of the group $U(1)\times U(1)$ over $\Sigma$.

Notice that the integral in condition \eqref{mn2}, also corresponds 
to a realization
of the central charge of the supersymmetric algebra of the
supermembrane. Hence the condition $n\not =0$ is
equivalent to having nontrivial central charge. If the $X_r$ fulfill condition
\eqref{mn1}, then \eqref{mn2} holds automatically. However non-triviality of
\eqref{mn1}, does not necessarily imply non-triviality of \eqref{mn2}.
We show below, that the configuration space of the compactified supermembrane
with fixed central charge is completely characterized by the integer $n$, only
the determinant of $(m_{jr})$ is relevant.

In order to describe the winding of the supermembrane in terms of maps
from $\Sigma$ onto $S^1\times S^1$ satisfying \eqref{mn2}, together with
maps \linebreak $(X^m)_{m=1}^7:\Sigma \longrightarrow \R^7$,
we interpret \eqref{mn2}
in terms of geometrical objects. Let
\begin{equation} \label{n5}
F:= \ud X_r \wedge \ud X_s\epsilon ^{rs}
\end{equation}
be a closed two-form on $\Sigma$, such that $F$ satisfies \eqref{mn2}.
Since $\ud F =0$ and $\int_\Sigma F=2\pi n$, there always exists a $U(1)$
principal
bundle over $\Sigma$ and a one-form connection on it, such that
$F$ is the curvature two-form, \cite{kost}. The integer $n$
characterizes the bundle.  If $n\not=0$, the bundle is non-trivial
and the connection one-form  must have non-trivial transitions.
In this bundle, there are particular connection one-forms satisfying
\begin{equation} \label{n7}
  ^\ast \widehat{F} \equiv \frac{\epsilon^{ab}}{\sqrt{W}} \widehat{F}_{ab}=n
\end{equation}
at any point of $\Sigma$, where $a,b=1,2$ are indices associated
to local coordinates on $\Sigma$. These are the so called Dirac
monopoles over Riemann surfaces. It turns out that these monopoles
together with the constraint $X^m=0$, $m=1,\ldots,7$, are
configurations where the Hamiltonian of the supermembrane have
local minima, \cite{8}. Moreover, there is only one
local minimum for each $n$.

Condition \eqref{n7} implies that
$\widehat{F}_{ab}$ is non-degenerate. Any non-degenerate closed
two-form, can always be decomposed as
\begin{equation*} \label{n8}
  \widehat{F}_{ab}=\partial_a\widehat{X}_r \partial_b \widehat{X}_s
   \epsilon^{rs},
\end{equation*}
in the sense that there exists a Darboux atlas for $\Sigma$ such that
the above holds on each open set. The $\widehat{X}_r$, $r=1,2$ are harmonic
maps over $\Sigma$ with metric
\[
   g_{ab}=\partial_a \widehat{X}_r \partial_b \widehat{X}_s \delta^{rs}
\]
the pull-back of the Euclidean metric over $S^1\times S^1$. Notice
that this is the metric arising from the supermembrane action. The
$\ud \widehat{X}_r$ are harmonic one-forms over $\Sigma$. If
$\Sigma$ is any given torus, we may then consider these one forms as
a normalized basis of homology over $\Sigma$. Clearly the maps
$\widehat{X}_r$ are not uniquely determined. In fact, we are
allowed to change the basis by an element of $SL(2,\Z)$, the
modular group, and $\widehat{F}$ remains invariant. This change of
basis corresponds to a conformal diffeomorphism over $\Sigma$.
From \eqref{n7} we obtain
\[
   n \sqrt{W}=\epsilon^{ab}\widehat{F}_{ab}=\epsilon^{ab}\partial_a
   \widehat{X}_r\partial_b \widehat{X}_s \epsilon^{rs},
\]
which is invariant under this conformal diffeomorphism.
Consequently it is an area preserving diffeomorphism and hence it
corresponds to a gauge symmetry of the supermembrane action.

Although $\widehat{X}_r$, $r=1,2,$ are not unique as homotopic
maps from $\Sigma$ onto $S^1\times S^1$, they are all equivalent
on the configuration space of the compactified supermembrane. Then
one realizes that the problem of handling the closed, but not
exact, one-forms in the quantum analysis of the compactified
$M_9\times S^1\times S^1$ supermembrane has been solved. In fact
the maps $X_r$ decompose as the sum of $\widehat{X}_r$ plus a
homotopically trivial map to be quantized. The $\widehat{X}_r$
will then be conveniently incorporated to the general description
of the action. In so doing, we will end up with a formulation of
the theory as  a symplectic non-commutative Yang-Mills action.

The zwei-vein
\begin{equation*} \label{n9}
   e^a_r \equiv \frac{\widehat{\Pi}^a_r}{\sqrt{W}} \equiv
   \frac{\epsilon^{ab}}{\sqrt{W}} \partial _b\widehat{X}_r,
\end{equation*}
allows us to write down all the geometrical objects in the spatial
world volume in terms of the corresponding objects in the tangent
space. We can express the curvature $F$ of any connection over the
mentioned bundle (characterized by $n$) in terms of $\widehat{F}$,
as
\begin{equation} \label{n11}
  F=\widehat{F} +f,
\end{equation}
where $f=\ud a$ satisfies $\int_\Sigma f=0$. The one-form $a$ is a
one-form connection on a trivial $U(1)$ bundle, it has no transitions over
$\Sigma$ and $f$ is an exact two-form.

Let
\begin{equation*}
   \mathrm{D}_r:= \frac{\widehat{\Pi}^a_r}{\sqrt{W}} \partial_a.
\end{equation*}
Let $\mathcal{A}_r$ be such that
\begin{equation*}
X_r=\widehat{X}_r+\mathcal{A}_r.
\end{equation*}
Then, by computing $\ud X_1 \wedge \ud X_2$, the decomposition \eqref{n11}
for \eqref{n5} yields
\begin{equation*} \label{n14}
  {}^\ast \!F= {}^\ast\!\widehat{F} + {}^\ast\!\mathcal{F},
\end{equation*}
where
\begin{gather*} \label{n15}
 {}^\ast\!\mathcal{F}= \epsilon^{rs} \mathcal{F}_{rs}, \\
 \mathcal{F}_{rs} = \mathrm{D}_r \mathcal{A}_s - \mathrm{D}_s \mathcal{A}_r
 +\{\mathcal{A}_r,\mathcal{A}_s\}.
\end{gather*}
It turns out, \cite{8}, that under the area preserving diffeomorphism,
the residual gauge symmetry of the supermembrane Hamiltonian in the
light cone gauge, $\mathcal{A}_r$, transforms as
\begin{equation*} \label{n16}
  \delta \mathcal{A}_r= \mathcal{D}_r \xi
\end{equation*}
where the covariant derivative
\begin{equation*} \label{n17}
  \mathcal{D}_r \cdot=\mathrm{D}_r \cdot +\{\mathcal{A}_r,\cdot\}.
\end{equation*}
Then, the term $\mathcal{F}_{rs}$ is interpreted as the curvature
of a symplectic non-commutative connection. The results of \cite{10}
describe the relationship between
this connection and the ones arising from a non-commutative product
on the Weyl algebra bundle.

The condition $\int _\Sigma f =0$ yields
\begin{equation} \label{n18}
  \int _\Sigma \mathcal{F} =0.
\end{equation}
This allows to write the Hamiltonian of the supermembrane, only in terms
of $\mathcal{A}_r$ and $X^m$. Identity \eqref{n18} arises by imposing
fixed central charge, or, analogously, by considering a
fixed  $U(1)$ principal bundle on $\Sigma$. Hence, according to
\cite{8},
\begin{equation}\label{e5}
\begin{aligned}
 H=\int_{\Sigma}&(1/2\sqrt{W})[ (P_{m})^{2}+(\Pi_{r})^{2}+
 (1/2)W\{X^{m},X^{n}\}^{2}
 +W(\mathcal{D}_r X^{m})^{2}+\\ & +(1/2)W(\mathcal{F}_{rs})^{2}]+
 \int_{\Sigma}[(1/8)\sqrt{W}n^{2}
 -\Lambda(\mathcal{D}_{r}\Pi_{r}+\{X^{m},P_{m}\})]+\\&
 -(1/4)\int_{\Sigma}\sqrt{W}n^{*}\mathcal{F}, \qquad \qquad
 n\not=0
\end{aligned}
\end{equation}
together with its fermionic contribution
\begin{equation}\label{e6}
\int_{\Sigma} \sqrt{W} [- \overline{\Psi}\Gamma_{-} \Gamma_{r}
\mathcal{D}_{r}\Psi +
 \overline{\Psi}\Gamma_{-} \Gamma_{m}\{X^{m},\Psi\} +
 \Lambda \{ \overline{\Psi}\Gamma_{-},\Psi\}].
\end{equation}
Here $P_m$ and $\Pi_r$ denote the momenta conjugate to $X^m$ and  $\CA_r$
respectively. By $\Psi$ we denote the Majorana spinors
of the $D=11$ formulation
which may be decomposed in terms of a complex 8-component spinor
of $SO(7)\times U(1)$.

The above Hamiltonian describes a non-commutative Maxwell
connection coupled to the transverse scalar fields to the
supermembrane world volume. The first class constraint generating
the area preserving diffeomorphisms realizes as the noncommutative
Gauss constraint. The presence of the integral of the
noncommutative curvature is highly relevant, since it explains  
 why  the nature of the 
spectrum in our model differs  from the  model studied 
in \cite{2}, \cite{2.1} and \cite{3}. Indeed, 
we can use the very same formulation \eqref{e5} of the Hamiltonian, 
which is also valid 
for the free winding supermembrane. We emphasize that
this is the exact expression for the Hamiltonian and not an approximation. 
In the free winding  case, $\mathcal{A}_r$ is a multi-valued connection over
$\Sigma$, unlike the fixed central charge case, where it is a single-valued 
object.
In this sense, our model corresponds to a restriction in the space of
all possible configurations of the free winding case.
In the fixed central charge model, the term $\sqrt{W}n^{*}\mathcal{F}$ 
is a total derivative, 
hence its integral cancels out. In this case the condition
of zero hamiltonian density in an open region implies both,
zero curvature and hence trivial $\mathcal{A}_r$, and  
constant $X^m$. This ensures the absence of singular configurations.
These results were obtained in \cite{1} in explicit manner. 
On the other hand, when $\mathcal{A}_r$ is multi-valued, the latter
term in expression \eqref{e5} can be non-zero and it is not difficult to
check that non-trivial singular configurations can arise in general. This
is in agreement with the results of \cite{3}.

Reference \cite{1} is devoted to finding an $SU(N)$ regularized
formulation of the Hamiltonian for a fixed non-trivial central charge.
In the case of free winding, the closed, but not exact, modes seem not to
fit in an $SU(N)$ model, cf. \cite{3}. Unfortunately, the present work 
does not contribute  
to this situation. However, as we mentioned earlier, in the 
case of fixed central charge,
the close one forms are given in unique manner in terms of a given basis 
of homology. The configuration space is then described in terms of exact
one-forms. The regularization procedure leads then to a model
formulated in terms of $SU(N)$-valued geometrical objects.
The resulting  model for a toroidal supermembrane is given by
\begin{align}
H= & \mathrm{Tr}\left(\frac{1}{2N^{3}}(P^{0}_mT_{0}P^{0}_{m}T_{0}+
\Pi_r^0T_{0}\Pi^{-0}_{r}T_{0}+(P_{m})^2+ (\Pi_{r})^{2})+
\right. \nonumber \\& +\frac{n^2}{16\pi^2N^3}[X^{m},X^{n}]^2+
\frac{n^2}{8\pi^2N^3}\left(\frac{i}{N}[T_{V_{r}},X^{m}]T_{-V_{r}}-
[\mathcal{A}_r,X^{m}]\right)^2+\nonumber \\&
+\frac{n^2}{16\pi^2N^3}\left([\mathcal{A}_r,\mathcal{A}_s]+
\frac{i}{N}([T_{V_s},\mathcal{A}_r]T_{-V_s}-[T_{V_r},\mathcal{A}_s]
T_{-V_r})\right)^2 + \frac{1}{8}n^2+ \nonumber \\& +\frac{n}{4\pi N^3}
  \Lambda\left([ X^{m},P_{m}]- \frac{i}{N}[T_{V_r},\Pi_{r}]T_{-V_r}
  +[ \mathcal{A}_{r},\Pi_{r}]\right)+ \nonumber \\
   &+ \frac{in}{4\pi N^3}(\overline{\Psi}\Gamma_{-}\Gamma_{m}
   \lbrack{X^{m},\Psi}\rbrack
   -\overline{\Psi}\Gamma_{-}\Gamma_{r}\lbrack{\mathcal{A}_{r},\Psi}
   \rbrack +
  \Lambda \lbrack{\overline{\Psi}\Gamma_{-},\Psi}\rbrack + \nonumber \\
  & - \left.\frac{i}{N} \overline{\Psi}\Gamma_{-}\Gamma_{r}
  [T_{V_{r}},\Psi] T_{-V_r})\right)  \label{e12}
\end{align}
subject to
\begin{align}
\mathcal{A}_{1}= &\mathcal{A}^{(a_1,0)}_{1}T_{(a_1,0)}, \nonumber \\
\mathcal{A}_{2}= &\mathcal{A}^{(a_1,a_{2})}_{2} T_{(a_1,a_2)}
\quad \textrm{with}\quad a_2\neq0. \label{e13}
\end{align}
Here $A=(a_1,a_2)$, where the indices $a_1,\,a_2=0,\ldots,N-1$ exclude
the pair $(0,0)$, $V_1=(0,1)$, $V_2=(1,0)$ and $T_0\equiv T_{(0,0)}=N\,
\mathbb{I}$. We agree in the following convention
\begin{equation*}
\begin{gathered}
 X^{m}=  X^{mA}T_{A},\quad\qquad P_{m}=P^{A}_m T_{A},\\
 \mathcal{A}_r= \mathcal{A}^A_r T_{A}, \quad\qquad
 \Pi_{r}=\Pi_r^{A}T_{A},
\end{gathered}
\end{equation*}
where $T_A$ are the generators of the $SU(N)$ algebra:
\begin{equation*}
 [T_{A},T_{B}]= f^{C}_{AB}T_{C}.
\end{equation*}

The condition (\ref{e13}) correspond to a truncation of a gauge
fixing of (\ref{e5}) and (\ref{e6}), when the geometrical objects are given
in a complete orthonormal basis of $L^{2}(\Sigma)$. Gauge fixing
conditions of the same kind were first used in \cite{2}.

We showed in \cite{1} (see also \cite{12}) that the regularized
Hamiltonian (\ref{e12}) has an associated mass operator with no
string-like spikes. The gauge fixing conditions
together with the constraint
\begin{equation*}
\begin{aligned}
 ([X^{m},P_{m}])=-\frac{i}{N}[T_{v},\Pi_{r}]T_{v}+
 [\mathcal{A}_{r},\Pi_{r}]+[\overline{\Psi}\Gamma_{-},\Psi]=0
 \end{aligned}
\end{equation*}
allow, \cite{12}, a canonical reduction of the Hamiltonian. Therefore the
conjugate pairs
\begin{equation*}
\begin{aligned}
& \mathcal{A}^{a,b}_{1},\Pi^{a,b}_{1},\qquad b\neq 0\\ &
\mathcal{A}^{a,0}_{2},\Pi^{a,0}_{2},
\end{aligned}
\end{equation*}
do not appear in (\ref{e12}). After this reduction
the terms $|\Pi_{1}^{a,b}|^{2}$, $b\neq 0$ and
$|\Pi_{2}^{a,0}|^{2}$ become nontrivial, however, since they are
positive, we can bound the mass operator by an operator $\mu$
without such terms. Notice that the imposition of the gauge fixing conditions
(\ref{e13}) together with the elimination of the associated conjugate
momenta, is valid only in the interior of an open cone $K$. Our
model considers then wavefunctions with support on the interior of
$K$. In order to show that the spectrum of $\mu$ is discrete, we
will consider $\mu$ as an operator acting on the whole
configuration space. The discreteness of the spectrum for the
latter implies the same property for the restriction to any
hyper-cone. We notice that the assumption that the quantum problem
is formulated only on an open cone $K$ of all configuration space,
with Dirichlet boundary conditions, is also implicit in
\cite{2}.

According to the results of \cite{12}, the bosonic part of the operator
$\mu$ has compact resolvent. Moreover, the bosonic
potential is what we can formally call ``basin shaped''.
We will prove in the forthcoming
sections that the operator $\mu$ including the fermionic sector,
has compact resolvent and consequently the mass operator of
the Hamiltonian (\ref{e12}) has also a compact resolvent.

\medskip


\section{The Fermionic Potential}
We consider the metric
\begin{equation*}
\begin{aligned}
\eta_{ab}=
 \begin{pmatrix}
    0&-1  \\
    -1&0\\
    \quad &\quad&1\\
     \quad&\quad&\quad &\ddots \\
    \quad & \quad & \quad& \quad&1
  \end{pmatrix}
\end{aligned}
\end{equation*}
in the light cone.
In the $2+2+7$ decomposition of the space-time indices,
$\Gamma^{\mu}=(\Gamma^{+},\Gamma^{-},\Gamma^{1},\Gamma^{2},\Gamma^{i})$
where
\begin{equation*}
(\Gamma^{+})^{2}=(\Gamma^{-})^{2}=0\qquad  \mathrm{and}
\qquad  \{\Gamma^{+},\Gamma^{-}\}=2\mathbb{I}.
\end{equation*}
An appropriate representation is found by considering
\begin{equation*}
\Gamma^{\pm}=\sigma_{\pm}\otimes \mathbb{I}\otimes \mathbb{I}_{8\times 8},
\end{equation*}
where
\begin{equation*}
\begin{aligned}
 \sigma_{\pm}&=\sqrt{\frac{1}{2}}(\sigma_{1}\pm i\sigma_{2}),\\
\Gamma^{r}&=\sigma_{3}\otimes i\sigma_{r}\otimes \mathbb{I}_{8\times
8},\qquad r=1,2,\\
\Gamma^{i}&=\sigma_{3}\otimes\sigma_{3}\otimes\gamma^{i}_{8\times
8}\qquad, i=3,..9
\end{aligned}
\end{equation*}
and
\begin{equation*}
(\gamma^{i})^{T}=-\gamma^{i},
\end{equation*}
in order to ensure that $C\Gamma^{i}$ are symmetric. The charge conjugation
is antisymetric and it is given by
\begin{equation*}
C=\sigma_{2}\otimes\sigma_{1}\otimes \mathbb{I}_{8\times 8}.
\end{equation*}
The Majorana condition and the light cone gauge fixing condition
associated to the kappa symmetry
\begin{equation*}
\overline{\Psi}=\Psi^{T}C, \qquad \mathrm{and} \qquad \Gamma^{+}\Psi=0,
\end{equation*}
respectively, allow to rewrite $\Psi$ in terms of a complex spinor
of $SO(7)$
\begin{equation*}
\Psi=
  \begin{pmatrix}
    i\chi^ {*}  \\
    \chi\\ 0\\0
  \end{pmatrix}
\end{equation*}
where $\chi_{\alpha}$ and $\chi_{\beta}^{+}$ are anticommuting
variables, and $\alpha,\beta=1,\ldots,8$ are $SO(7)$ spin indices.
The canonical quantization rules for  $\chi$, considering the
second class
Dirac constraint on the spinor fields, turns out to be
\begin{equation} \label{e1}
\begin{aligned}
&\{\chi_{\alpha},\chi_{\beta}^{*}\}=\delta_{\alpha\beta}\\
&\{\chi_{\alpha},\chi_{\beta}\}=\{\chi_{\alpha}^{*},\chi_{\beta}^{*}\}=0.
\end{aligned}
\end{equation}
In terms of the matrix model, the canonical quantization
reads
\begin{equation*}
\begin{aligned}
\{\chi_{\alpha}^{A},\chi_{\beta
B}^{\dag}\}&=\delta_{\alpha\beta}\delta_{B}^{A}\\
\{\chi_{\alpha}^{A},\chi_{\beta}^{B}\}&=\{\chi_{\alpha A
}^{\dag},\chi_{\beta B}^{\dag}\}=0.
\end{aligned}
\end{equation*}

This allows us to compute the fermionic potential in terms of the $\chi$
fields. We have
\begin{equation*}
\langle
\overline{\Psi}\Gamma^{-}\Gamma_{m}X^{m}\partial\Psi\rangle
=\langle
-\sqrt{2}(-i\chi^{\dag}\gamma_{m}\{X^{m},\chi\}+i\chi^{\dag}
\gamma_{m}\{X^{m},\chi^{*}
\})\rangle
\end{equation*}
and in the matrix model
\begin{equation*}
\begin{aligned}
\frac{-n}{4\pi N^{3}}f_{BC}^{A}
[\chi^{\dag}_{A}\gamma_{m}X^{mB}\chi^{C}+\chi^{TC}
\gamma_{m}X^{mB}\chi^{*}_{A}].
\end{aligned}
\end{equation*}
The other terms are constructed in a similar manner. Spinors will
be represented by $2^{8D}\times 2^{8D}$ matrices, with
$D$ equals to the dimension of the symmetry group, in this case $SU(N)$.
The explicit
construction
of such matrices
can be performed inductively. To avoid dispersing from our
main task, we show this procedure in detail in section~6.

\section{``Basin shaped'' potentials}
The criterion we present below extends the well-known fact that if
a potential is bounded from below and $V(x)\to +\infty$ as $|x|\to
\infty$, then the Hamiltonian $-d^2/dx^2+V$ as a closed operator
acting on $L^2(\R^n)$ has discrete spectrum (see for instance
\cite[theorem XIII.16]{13.2}). For completeness of our exposition,
we provide a self-contained proof of this result.

The Hamiltonian we shall consider is as follows. Let
\[
   H= -\lap + V
\]
acting as a closed operator on $L^2(\R^N)\otimes \C^n$, where
$\lap\equiv \lap_x\otimes \mathbb{I}_{n\times n}$ and
$V(x)=V(x)^\ast\in \C^{n\times n}$, $x\in \R^N$. In addition,
assume that $V$ is measurable and satisfies $V(x)\geq c$ in the
sense that
\[
    V(x)w\cdot w \geq c,\qquad x\in \R^N, w\in \C^n
\]
where $c\in \R$ is constant.
This ensures that the operator $H$ can be defined by means of a
symmetric quadratic form and it is bounded from below. Once we have defined
rigorously operator $H$, the validity of the following criterion is ensured.

\begin{lemma} \label{t1} Let $v_k(x)$ be the
eigenvalues of $V(x)$.
If all $v_{k}(x)\to +\infty$ as $|x|\to \infty$, then the spectrum
of $H$ is discrete.
\end{lemma}

\Proof Without loss of generality we assume that $v_k(x)\geq 0$.
Since $H$ is bounded from below, one can apply the
Raleigh-Ritz principle to find the eigenvalues below the essential
spectrum. Let
\[
  \lam_m(T):=\inf \left(\sup_{\Phi\in L}
  \frac{\dotp{T\Phi,\Phi}}{\|\Phi\|^2}\right)
\]
where the infimum is taken over all $m$-dimensional subspaces
$L\subset L^2(\R^N) \otimes \C^n$. Then the bottom of the
essential spectrum of $T$ is $\lim_{m\to \infty}\lam_m(T)$. Notice
that if this limit is $+\infty$, then the spectrum of $T$ is
discrete. Above
\[
  \dotp{\Phi,\Psi} = \int_{\R^n} \Phi(x) \cdot \Psi(x) \ud x.
\]
and $\|\cdot\|^2=\dotp{\cdot,\cdot}$.

The hypothesis of the lemma is equivalent to the following
condition: for all $c>0$, there exists a ball $S$ (of possibly
very large radius) such that
\[
 V(x)w\cdot w \geq c|w|^2,\qquad \mathrm{all}\ x\not \in S.
\]

Let
\[
 W(x):=\left\{ \begin{array}{ll} -c & x\in S \\ 0 & x
 \not \in S \end{array} \right. .
\]

Then for all $\Phi$ smooth and with compact support,
\[
  V(x)\Phi(x) \cdot \Phi(x) \geq c\Phi(x)\cdot \Phi(x) +
  W(x)\Phi(x) \cdot \Phi(x)
\]
so that
\[
  \dotp{V\Phi,\Phi} \geq \dotp{(c+W)\Phi,\Phi}.
\]
Thus
\[
 \lam_m(H) \geq c+ \lam_m(-\lap + W)
\]
for all $l=1,2,\ldots$.

Since $W(x)$ is a bounded potential with compact support, by
Weyl's theorem, the essential spectrum of $-\lap+W$ is
$[0,\infty)$. Therefore by Raleigh-Ritz criterion, there exists
$\tilde{M}>0$ such that
\[
   \lam_m(-\lap+W) \geq -1,\qquad m\geq \tilde{M}.
\]
Thus, $\lam_m(H)\geq c-1$ for all $m\geq M$. Since we can take $c$
very large, necessarily $\lam_m\to +\infty$ as $m$ increases and
the proof of the lemma is complete.


\section{Discreteness of the spectrum}
By using lemma \ref{t1}, we show in this section
that the resolvent of the operator $\mu$ is compact and hence it has discrete spectrum.

To this end, decompose
$\mu$ as
\begin{equation*}
\mu=-\Delta+V_{B}\mathbb{I}+V_{F}
\end{equation*}
where $V_B$ and $V_{F}$ denote the bosonic
and fermionic
potentials respectively. Then $V_{F}$ is the sum of a linear homogeneous part
$M(X,\mathcal{A})$ corresponding to
\begin{equation*}
   \frac{in}{4\pi N^3}(\overline{\Psi}\Gamma_{-}\Gamma_{m}
   \lbrack{X^{m},\Psi}\rbrack
   -\overline{\Psi}\Gamma_{-}\Gamma_{r}\lbrack{\mathcal{A}_{r},\Psi}
   \rbrack )
\end{equation*}
and a constant matrix $C$ corresponding to
\begin{equation*}
 \frac{-n}{4\pi N^4}(\overline{\Psi}\Gamma_{-}\Gamma_{r}
  [T_{V_{r}},\Psi] T_{-V_r}).
\end{equation*}

Put  $T:=-\Delta+V_{B}+M(X,\mathcal{A})$. Since $T$
is Hermitian (a self-adjoint operator in its domain), $(T-i\rho)$ is invertible
for all $\rho>0$ and we can choose $\rho$
large enough such that the resolvent $(T-i\rho)^{-1}$ satisfies
\[
   \|(T-i\rho)^{-1}\|\leq \|C\|^{-1}/2.
\]
Here $\|\cdot\|$ is the supremum norm for operators acting on Hilbert spaces.
Hence
\[
  \mu-i \rho = T+C -i\rho =(T-i\rho)(\mathbb{I}+(T-i\rho)^{-1}C).
\]
Because of $\|(T-i\rho)^{-1}C\|\leq 1/2$, the latter term is invertible and so
\[
  (\mu-i \rho)^{-1} = (\mathbb{I}+(T-i\rho)^{-1}C)^{-1}(T-i\rho)^{-1}.
\]
Since the first term at the right hand side is bounded, the
resolvent of $\mu$ is compact if and only if the resolvent of $T$
is compact.

We apply lemma \ref{t1} in order to show compactness for the resolvent of $T$.
If we denote by  $R$ the normal vectors in the configuration space so that $X=R\phi$ and
$\mathcal{A}=R\psi$, according to \cite[\S 3]{12}, $R=0$ is a double zero
of $V_B(R\phi,R\psi)$  and
\begin{equation*}
V_{B}(R\phi,R\psi) \geq k R^{2}
\end{equation*}
for some constant $k>0$. The eigenvalues of the matrix
\begin{equation*}
V_{B}+M(X,\mathcal{A})
\end{equation*}
are the $\lambda\in \R$ such that
\begin{equation*}
\det [\lambda-V_{B}-M(X,\mathcal{A})]=0.
\end{equation*}
By virtue of the homogeneity of $M$, $\lam$ must satisfy
\begin{equation*}
\det \left[
\frac{\lambda-V_{B}}{R}\mathbb{I}-M(\phi,\psi)\right]=0, \qquad R>0.
\end{equation*}
Therefore if $\widehat{\lambda}$ are the eigenvalues of
$M(\phi,\psi)$, then
\begin{equation*}
\lambda=V_{B}(R\phi,R\psi)+R\widehat{\lambda}.
\end{equation*}
Consequently, $\lambda\to +\infty$ whenever $R\to\infty$. Notice
that $V$ is continuous, hence it is automatically bounded from below.
This ensures that the resolvent of $T$ is compact as a consequence of
 lemma \ref{t1}.

\section{Matrix representation of spinors}
In this final section we show how to construct a basis
$\chi_\alp,\,$ \linebreak $\alp=0,\ldots,n-1$ of
size $2^n\times 2^n$  satisfying the anti-commutative relations
\begin{equation} \label{e17}
\begin{aligned}
  \{\chi_\alpha,\chi_\beta^{\ast}\}&=\delta_{\alpha \beta}, \\
  \{\chi_\alpha,\chi_\beta\}&=0.
\end{aligned}
\end{equation}
This ensures \eqref{e1}.

The entries of the matrices $\chi_\alpha$ are
either $0$ or $\pm 1$. Adopting a common notation in
combinatorics, the list
\[
    [(m_1,n_1)_\pm;\ldots;(m_j,n_j)_\pm]
\]
denotes a matrix full of zeros except that it has either $+1$ or $-1$ at the
entries $(m_1,n_1),\ldots,(m_j,n_j)$. For $l=0,1,\ldots$ and $\alpha>0$, let
\begin{align*}
   B_0(l)&=(2l+1,2l+2)_+ \\
   B_\alpha(l)&=[(2^{\alpha+1}l\!+\!1,2^{\alpha+1}l\!+\!1\!+\!2^\alpha)_s;\ldots;
   (2^{\alpha+1}l\!+\!(2m\!+\!1),2^{\alpha+1}l\!+\!(2m\!+\!1)\!+\!2^\alpha)_s; \\
   &\ldots;(2^{\alpha+1}l\!+\!(2^\alpha\!-\!1),2^{\alpha+1}l\!+\!
   (2^\alpha\!-\!1)\!+\!2^\alpha)_s;
   (2^{\alpha+1}l\!+\!2,2^{\alpha+1}l\!+\!2\!+\!2^\alpha)_s; \\
   &\ldots;(2^{\alpha+1}l\!+\!(2m\!+\!2),2^{\alpha+1}l\!+\!(2m\!+\!2)\!+\!2^\alpha)_s;
   \ldots
     ;(2^{\alpha+1}l\!+\!2^\alpha,2^{\alpha+1}l\!+\!2^\alpha\!+\!2^\alpha)_s]
\end{align*}
be blocks of size $2^{\alpha+1}\times 2^{\alpha+1}$, where the sign $s$
of the
non-zero entry $(m,n)_s$ of $B_\alpha(0)$ is determined according to the rule
\begin{equation} \label{e111}
  s=(-1)^{\#(q)_2}(-1)^{m+1}, \qquad m=2q+1\quad
  \mathrm{or} \quad m=2q+2
\end{equation}
and the sign distribution of $B_\alpha(l)$ for $l\geq 1$ is the same as
that of $B_\alpha(0)$. The integer $\#(q)_2$ is the number of ones in
the binary representation of $q$. Then we construct the desired
basis as the block diagonal matrices
\[
   \chi_\alpha=\left[B_\alpha(0);\ldots;B_\alpha(2^n/2^{\alpha+1}-1)\right],
\qquad \alpha=0,\ldots,n-1
\]
of size $2^n\times 2^n$. In order to illustrate this procedure
take $n=4$, then the basis is {\small
\begin{gather*}
  \chi_0=[(1,2)_+;(3,4)_+;(5,6)_+;(7,8)_+;(9,10)_+;(11,12)_+;
  (13,14)_+;(15,16)_+]
  \\
  \chi_1=[[(1,3)_+;(2,4)_-];[(5,7)_+;(6,8)_-];[(9,11)_+;(10,12)_-];
  [(13,15)_+;(14,16)_-]]
  \\
  \chi_2=[[(1,5)_+;(3,7)_-;(2,6)_-;(4,8)_+];[(9,13)_+;(11,15)_-;
  (10,14)_-;(12,16)_+]]
  \\
  \chi_3=[(1,9)_+;(3,11)_-;(5,13)_-;(7,15)_+;(2,10)_-;(4,12)_+;
  (6,14)_+;(8,16)_-].
\end{gather*}}

We now show that the $\chi_0,\ldots,\chi_{n-1}$ satisfy
\eqref{e17}. In order to simplify notation, we
represent the product of two matrices whose rows and columns have
only one non-zero entry whose value is $\pm 1$, as a signed
permutation of the group of integer numbers. To be more precise,
it is easy to see that the only non-zero entries of this product
are the pairs $(m,o)_s$ with sign $s=rt$ where the entries
$(m,n)_t$ are non-zero with sign $t$ on the first matrix, and the
entries $(n,o)_r$ are non-zero with sign $r$ on the second matrix.
This will be described pictorially as $m\stackrel{t}{\rightarrow}
  n\stackrel{r}{\rightarrow} o.$

\underline{$\{\chi_\alpha,\chi_\alpha\}=0$ and
$\{\chi_\alpha,\chi_\alpha^\ast\}=\mathbb{I}$}. Notice that each
\[
  \chi_\alpha=(m_1,n_1)_s;\ldots;(m_j,n_j)_s,
\]
where each index $m_p$ or $n_p$ is different from all other
indices. This ensures the first identity. The second identity is
consequence of the fact that the only non-zero entries of the
product $\chi_\alpha \chi_\alpha^\ast$ are represented by
\[ m_p\stackrel{s}{\rightarrow}
n_p\stackrel{s}{\rightarrow} m_p\] and those of
$\chi_\alpha^\ast\chi_\alpha$ by \[n_p\stackrel{s}{\rightarrow} m_p
\stackrel{s}{\rightarrow} n_p.\]

\medskip

In order to show that $\{\chi_\alpha,\chi_\beta\}=0$ when $\alpha\not=\beta$,
it is enough
to check this property only when $0\leq \alpha<n-1$ and $\beta=n-1$, whereas the
other cases follow from an inductive argument.
Furthermore since our basis consists of block diagonal
matrices, we only have to verify how the first block of $\chi_\alpha$, $B_\alpha(0)$,
multiplies with the suitable elements of $\chi_\beta$. Notice that the
only possibly non-zero entries of $\{\chi_\alpha,\chi_\beta\}$ are
$(m,n)$ consequence of the sum
\begin{equation} \label{e18}
  m\stackrel{s}{\rightarrow} o
\stackrel{t}{\rightarrow} n\quad +\quad m\stackrel{u}{\rightarrow}
p \stackrel{v}{\rightarrow} n.
\end{equation}
We claim that this entry will also be zero, due to the fact that the product of
signs $stuv$ is always negative. Let us show this claim.

\underline{$\{\chi_0,\chi_{n-1}\}=0$}. Put \[m=1,\, o=2,\,
n=2+2^{n-1},\, p=1+2^{n-1}\] in \eqref{e18}. The signs $s$ and $v$ are
positive because of all the non-zero entries of $\chi_0$ are equal
to one. According to the rule (\ref{e111}),
$t=(-1)^{\#(0)_2}(-1)^2$ and $u=(-1)^{\#(0)_2}(-1)^3$. This
ensures that the product of signs is always negative in this case.

\underline{$\{\chi_\alpha,\chi_{n-1}\}=0,\, 0<\alpha<n-1$}. If $m$ is
odd, the indices in \eqref{e18} are
\[m\!=\!2q\!+\!1,\, o\!=\!2q\!+\!1\!+\!2^\alpha,\,
n\!=\!2q\!+\!1\!+\!2^\alpha\!+\!2^{n-1},\, p\!=\!2q\!+\!1\!+\!2^{n-1}\]
for $0\leq q\leq 2^{\alpha-1}-1$. Then
\[
  s=u=(-1)^{\#(q)_2}=v,\qquad t=(-1)^{\#(q+2^{\alpha-1})_2}.
\]
For the third inequality notice that by construction the sign rule
of $B_\alpha(l)$ for $l>0$ copies the one of $B_\alpha(0).$ Since
the binary representation of $q$ has at most $2^{\alpha-2}$ digits,
then $\#(q+2^{\alpha-1})_2=\#(q)_2+1$ and thus the product of signs is
always negative. If $m$ is even, for reasons similar to the odd
case, also $s=u=v$ and $t=-s$.

This completes the proof of our claim.

\medskip

\underline{$\{\chi_\alpha,\chi_\beta^\ast\}=0,\, \alpha\not=\beta$}.  By using
arguments involving the diagram \eqref{e18}, this can be
shown in a similar manner as the previous identity. One should take into
account that due to the transposition in the second term, the
signs of $s$ and $v$ should be determined not from the first
but from the second entry of the suitable pair.


\section*{Conclusions}
We have shown that the quantum Hamiltonian of the compactified
supermembrane on $M_9\times S_1\times S_1$ with non-trivial
central charge, that is with irreducible winding, has compact resolvent and
consequently its spectrum consists of a discrete set of
eigenvalues with finite multiplicity.

We have considered a regularized $SU(N)$ model of the compactified
supermembrane with non-trivial central charge. The condition of
having a non-trivial winding, determines a sector of the full
compactified supermembrane. In the explicit formulas we use a
toroidal supermembrane, however, the result still holds for other
non-trivial topologies (the spherical supermembrane has been
considered recently in \cite{15} in terms of an $SU(2)$ model). Our
approach is valid for the analysis of the compactified
supermembrane with target space $M_9\times S_1\times S_1$. In this
case the class of maps defining the configuration space of the
supermembrane are determined by the central charge which becomes
proportional to $n$, the winding number of the supermembrane. The
existence of a non-trivial central charge leads to a
re-formulation of the problem in terms of a symplectic
non-commutative super Yang-Mills theory. In particular our
regularized Hamiltonian is a consequence of this construction.


The lemma we have established in section $5$, seems to be the
appropriate strategy to investigate any compactified
supersymmetric models where no string-like configurations are
present. The assumptions on the bosonic potential are very mild,
we only require the potential to be measurable, bounded from below
and unbounded above in every direction (if the potential is
continuous the unbounded assumption ensures the boundedness from
below). For instance, for a quantum mechanical potential of the form
\[
  V=V_B(x)\mathbb{I}+V_F(x) \in \C^{2^n\times 2^n},
\]
where $x\in \R^L$,
$V_B(x)$ is continuous with the asymptotic behaviour
\[
   V_B \geq c \|x\|^{2p}, \qquad c>0,
\]
and the fermionic matrix potential satisfies
\[
 V_F\leq V_F|_{\|x\|=1} \|x\|^q,
\]
for all $\|x\|>R_0$ with $2p>q$, the Hamiltonian
of the quantum system has spectrum consisting exclusively 
of isolated eigenvalues of finite multiplicity.

\medskip

{\scshape Acknowledgments.} We wish to thank M.~Rosas for helping
us with the combinatoric aspect of section 6. We also wish to
thank J.~Pe\~{n}a, M.~Asorey, T.~Ort\'\i n,  P.~Howe and K.~Stelle
for helpful discussions.


\begin{thebibliography}{99}
\bibitem{1}{\scshape M.P. Garc\'{\i}a del Moral, A. Restuccia},
``On the spectrum of a noncommutative formulation of the D=11 supermembrane
with winding''. {\it Phys. Rev. D} 66 (2002) 045023. \texttt{hep-th/0103261}.
\bibitem{2} {\scshape B. de Wit, M. L\"uscher, H. Nicolai}, ``The supermembrane
is unstable''. {\it Nucl. Phys. B} 320 (1989) 135-159.
\bibitem{2.1}{\scshape B. de Wit, Marquardt, H. Nicolai}, ``   ''
{\it Comm. Math. Phys.} 128 (1990) 39. 
\bibitem{3}{\scshape B. de Wit, K. Peeters, J.C. Plefka,} ``The supermembrane
with winding''. {\it Nucl. Phys. Proc. Suppl.} 62 (1998) 405-411.
\texttt{hep-th/9707261}.
\bibitem{4}{\scshape R. Helling, H. Nicolai}, ``Supermembranes and M(atrix)
theory''. Preprint 1998, AEI-093. \texttt{hep-th/9809103}.
\bibitem{5} {\scshape A. de Castro, A. Restuccia}, ``Master canonical action
and BRST charge of the M-theory bosonic five brane''. {\it Nucl. Phys. B} 617
(2001) 215-236. \texttt{hep-th/0103123}.
\bibitem{6} {\scshape I. Bandos, K. Lechner, A. Nurmagambetov, P.~Pasti,
D.~Sorokin, M.~Tonin}, ``Covariant action for the super-five-brane of
M-theory'' {\it Phys. Rev. Lett.} 78 (1997) 4332-4334. \texttt{hep-th/9701149}.
\bibitem{7} {\scshape M. Aganagic, J. Park, C. Popescu, J.~Schwarz},
``World-volume action of the M-theory five brane''
{\it Nuc. Phys. B} 496 (1997) 191-214. \texttt{hep-th/9701166}.
\bibitem{8}{\scshape I. Mart\'{\i}n, J. Ovalle, A. Restuccia},
``Stable solution of the double compactified $D=11$ supermembrane
dual''. {\it Phys. Lett. B} 472 (2000) 77-82. \texttt{hep-th/9909051}.
\bibitem{9}{\scshape I. Mart\'{\i}n, J. Ovalle, A. Restuccia},
``Compactified D=11 supermembranes and symplectic noncommutative
gauge theories''. {\it Phys. Rev. D} 64 (2001) 046001. \texttt{hep-th/0101236}.
\bibitem{10}{\scshape I. Mart\'{\i}n, A. Restuccia},
``Symplectic connections, noncommutative Yang Mills theory and
supermembranes''. {\it Nucl. Phys. B} 622 (2002) 240-256. \texttt{hep-th/0108046}.
 \bibitem{11}{\scshape J.G. Russo}, ``Supermembrane dynamics from
multiple interacting strings''. {\it Nucl.Phys.B} 492 (1997) 205-222.
\texttt{hep-th/9610018}.
\bibitem{12}{\scshape L. Boulton, M.P. Garc\'{\i}a del Moral, I.~Mart\'{\i}n,
A.~Restuccia}. "On the spectrum of a matrix model for the $D=11$ supermembrane
  compactified on a torus with non-trivial winding".
{\it Class. Quantum. Grav.} 19(2002) 2951-2959. \texttt{hep-th/0109153}.
\bibitem{13.1}{\scshape B. Simon},
``Some quantum operator with discrete spectrum but classically
continuous spectrum''. {\it Ann.Phys. NY} 146 (1983) 209-20.
\bibitem{13.2}{\scshape M. Reed, B. Simon}, {\em Methods of modern
mathematical physics, volume 4: analysis of operators}, Academic
press, New York, 1978.
\bibitem{kost} {\scshape B. Kostant}, ``Quantization and unitary representations. I.
Prequantization''. In {\em Lecture Notes in Math.}, Vol. 170,
Springer, Berlin, 1970, pp 87-208.
\bibitem{14}{\scshape M.J. Duff, T. Inami, C.N. Pope, E.~Sezgin, K.~Stelle.}
``Semiclassical quantization of the supermembrane''.
{\it Nucl. Phys. B} 297 (1988) 515.
\bibitem{15}{\scshape J. Conley, B. Geller, M.G. Jackson, L. Pomerance,
S. Shrivastava}.
``A quantum mechanical model of spherical supermembranes''. Preprint, 2002.
\texttt{hep-th/0210049}.
\end{thebibliography}
\end{document}